\documentclass[twocolumn,pre,showpacs,amsmath,amssymb]{revtex4}
\usepackage{hyperref}
\usepackage[pdftex]{graphicx}  
\usepackage{amssymb,amsfonts,amsmath}
\usepackage{dcolumn}   
\usepackage{bm}        
\usepackage{color}

\newcommand{\br}{\mathbf{r}}              
\newcommand{\mism}{\mathbf{d}}            
\newcommand{\Mism}{\mathbf{D}}            
\newcommand{\ave}[1]{\langle #1 \rangle}  
\newcommand{\sconf}{S_{\rm conf}}         

\begin{document}

\title{Probing a critical length scale at the glass transition}
\author{Majid Mosayebi}
\author{Emanuela Del Gado}
\author{Patrick Ilg}
\author{Hans Christian \"Ottinger}
\affiliation{Polymer Physics, ETH Z\"urich, Department of Materials, CH-8093 Z\"urich, Switzerland}
\date{\today}

\begin{abstract}
We give evidence of a clear structural signature of the glass transition, in terms of 
a static correlation length with the same dependence on the system size which is typical of critical 
phenomena. Our approach is to introduce an external, static perturbation to extract the structural 
information from the system's response. 
In particular, we consider the transformation behavior of the local minima of the underlying potential 
energy landscape (inherent structures), under a static deformation. The finite-size scaling analysis 
of our numerical results indicate that the correlation length diverges at a temperature $T_c$, below the 
temperatures where the system can be equilibrated.
Our numerical results are consistent with random first order theory, 
which predicts such a divergence with a critical exponent $\nu=2/3$
at the Kauzmann temperature, where the extrapolated configurational entropy vanishes.
\end{abstract}


\maketitle

Is there a static, structural origin of the dramatic slowing down of dynamics in supercooled liquids? 
The physical mechanism of the dramatic slowing down and corresponding increase in viscosity upon 
approaching the glass transition has been vehemently debated for more than 40 years 
\cite{kob-binder,Debenedetti_review}.
In critical phenomena, the slowing down of the system's dynamics is intimately related to a nearby 
phase transition, where a diverging length scale typically reflects the onset of long-range 
correlation. Several recent theories of the glass transition are built in this spirit 
\cite{wolynes89,MezardParisi2000,Procaccia_StatMechGlass,cavagna_pedestrian} 
and are supported from results on dielectric 
susceptibility \cite{Menon_divergesuscept} 
and specific heat \cite{Fernandez_specheat,Coluzzi00}. 
The few experimental studies, however, show only a moderate cooperativity length near the 
glass transition of some four to five particle diameters \cite{Donth_length}. 
Specifically designed simulations \cite{Biroli_overlapRFOT} have recently 
indicated a growing amorphous order in the range of four particle diameters. 
The results confirm, to some extent, the predictions of 
random first order theory (RFOT) \cite{wolynes89,cavagna_pedestrian}. 
However, no evidence for critical behavior with a diverging length scale 
could be found there. 
A growing correlation length is also indicated by a very recent numerical 
study of different local quantities \cite{Tanaka_critical} and it is interpreted 
as the hint of a critical behavior, whose evidence, nevertheless, is still very elusive.   
The heterogeneous dynamics of glass forming liquids allow one to define a dynamical correlation length 
\cite{kob-binder,Harrowell_Reichmann,Ludo_Science05,Franz,Andersen_review05} 
which is indeed growing as the glass transition is approached, but it remains 
unclear how, or whether at all, this dynamical length  
has a static, structural origin as is the case in phase transitions.  
A promising, alternative approach to supercooled liquids is based on inherent structures, 
i.e.~the local minima of the underlying potential energy landscape 
\cite{StillingerWeberIS2}. 
Although several qualitative changes in the system dynamics can be related to changes in 
their inherent structures \cite{SastryIS_fragile,XiaWolynes2000}, 
no length scale as significantly increasing as the dynamical correlation length has emerged so far
\cite{Sastry_growinglength}. 

Following a nonequilibrium thermodynamic theory of glasses \cite{hco_glass}, here 
we use small, static deformations to perturb the inherent structure 
configurations of supercooled liquids approaching the glass transition \cite{ema_mismatch}.
With this procedure, large correlated regions emerge at low temperatures and
allow us to detect a static correlation length that shows critical behavior
upon approaching the glass transition. 
%

{\em Methods and numerical simulation.} --  
We employ a binary Lennard-Jones mixture, which is an established model for 
fragile glass formers \cite{KobAndersen_1995a} (see Appendix \ref{app1}). 
An ensemble of well equilibrated configurations are 
prepared at constant density by slowly cooling statistically 
independent samples 
from high temperatures down to the supercooled regime. 
We study different system sizes ranging from $N=2000$ up to $N=64000$ particles. 
The inherent structure $X^{\rm q}=\{\br_j^{\rm q}\}$ corresponding to the 
actual configuration $X=\{\br_j\}$ is obtained by locally minimizing the system's 
potential energy by a conjugate gradient method. 
The potential energy landscape and its local minima (the inherent structures) 
have been studied intensively in recent years, 
unraveling a number of remarkable relations between inherent structure 
properties and the system's behavior  
\cite{Heuer_ISreview,Yip,Pastore_localsaddle,Ashwin,Doros_IS2}. 
Unlike these previous studies, 
we investigate the relation between two inherent structure configurations, 
$X^{\rm q}$ and $X^{\rm dq}$, where the latter is the inherent structure 
corresponding to the affinely deformed configuration 
$X^{\rm d}=\{\br_j^{\rm d}\}$, 
$\br_j^{\rm d}={\bf E}\cdot\br_j$, where the deformation is represented 
by the matrix ${\bf E}$. 
From these configurations, we define the mismatch vectors 
$\mism_j\equiv \br_j^{\rm dq}-{\bf E}\cdot\br_j^{\rm q}$, 
which give the nonaffine displacements between two inherent structure configurations. 
In particular, we consider static shear deformations with small amplitude $\gamma$, 
${\bf E}={\bf 1}+\gamma{\bf e}_1{\bf e}_2$, where ${\bf e}_\alpha$ are 
Cartesian unit vectors. 
In Ref.~\cite{ema_mismatch}, we found, as the temperature is lowered towards 
the glass transition, characteristic changes 
which are strongly reminiscent of the systems long time dynamics: 
small average mismatch lengths which are correlated over large distances.

\begin{figure}
\includegraphics[width=1.0\linewidth]{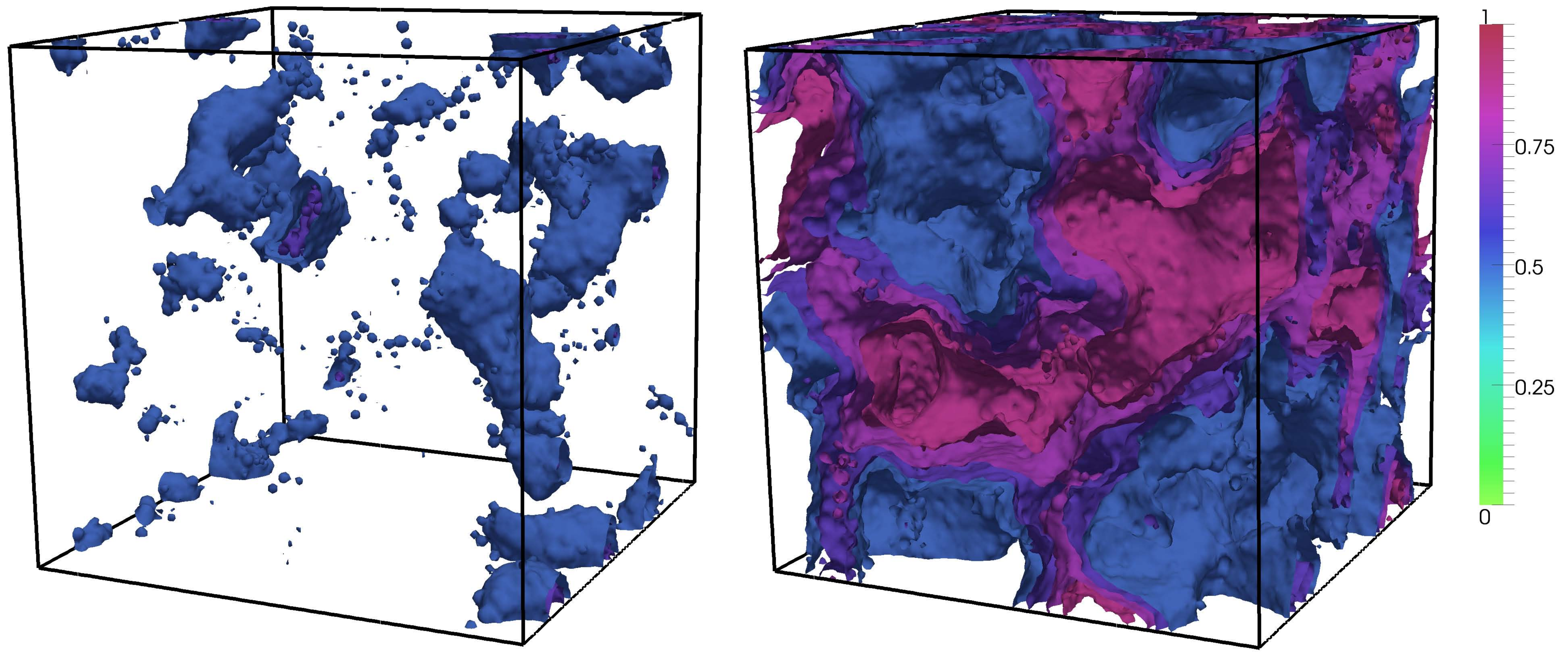}
\caption{(color online).
Coarse-grained nonaffine displacement field $\Mism_b(\br)$ defined 
in Eq.~(\ref{dCG}) for high ($T=1.0$, left panel) and low ($T=0.4$, right panel) temperature. 
The coarse-graining length was chosen as $b=2.0$.
The color code illustrates surfaces of constant $|\Mism|^2$-values. 
For better visibility, only particles with $|\Mism_j|^2\geq 0.5$ are shown.}
\label{Dfield.fig}
\end{figure}

{\em Results.} --  
In order to study correlations in the direction of the mismatch field, we define 
the coarse-grained nonaffine displacement field
$\Mism_b(\br)=\sum_j \Mism_j(b)\delta(\br-\br_j^{\rm q})$
of the inherent structure configuration $X^{\rm q}$, where $\Mism_j(b)$ are
the coarse-grained mismatch orientations obtained
by averaging $\mism_j'=\mism_j/|\mism_j|$ over a sphere of radius $b$ \cite{Fabien_prl}, 
\begin{equation} \label{dCG}
 \Mism_j(b) = N_j^{-1}\sum_{k}\mism_k'\chi_b(r_{jk}^{\rm dq}).
\end{equation}
Here, 
$r_{jk}^{\rm dq}$ is the distance between particles $j$ and $k$ in the 
inherent structure of deformed configuration 
$X^{\rm dq}$, $N_j = \sum_{k}\chi_b(r_{jk}^{\rm dq})$ 
the number of neighbors of particles $j$ within a distance $b$, and 
$\chi_b(r)=1$ if $r\leq b$ and zero elsewhere. 
For sufficiently small deformations, the mismatch field is 
approximately independent of strain amplitude $\gamma$ in a statistical sense 
and the use of normalized mismatch vectors 
$\mism_j'$ suppresses the strong temperature 
dependence of the average mismatch lengths~\cite{ema_mismatch}. 
 
In Fig.~\ref{Dfield.fig}, $\Mism_b(\br)$
of a typical configuration is shown at high and low temperature. 
The large domains where $|\Mism_j|^2\geq 0.5$ at low temperature clearly indicate large regions 
where the response to the quasi-static external deformation is strongly correlated, 
in marked contrast to the high temperature behavior.  

\begin{figure}
\includegraphics[width=1.0\linewidth]{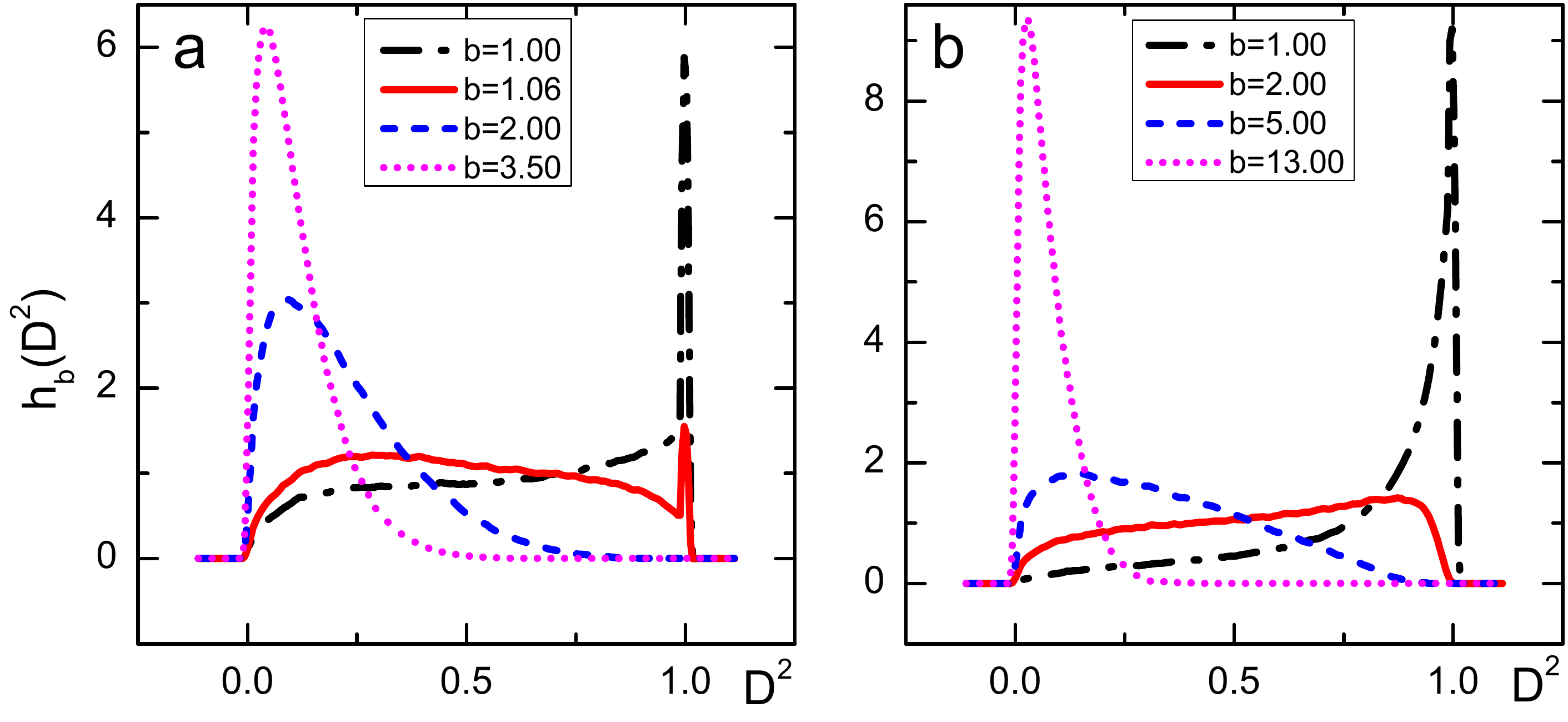}
\caption{Histogram of mismatch lengths $h_b(\Mism^2)$ at high (a: $T=1.0$) and low (b: $T=0.4$)
  temperature for different values of the coarse-graining length $b$. At high 
temperatures, the typical value of $\Mism^2$ jumps from an ordered 
($\Mism^2 \approx 1$) to a disordered one ($\Mism^2 \approx 0$) when $b$ 
becomes slightly larger than the particle diameter. 
At low temperatures, instead, this transition happens at much larger values of $b$.}
\label{hist_mismatch.fig}
\end{figure}

These observations can be made more quantitative by studying the distribution 
of coarse-grained mismatch lengths $h_b(\Mism^2)$. 
For small $b$, only a single particle contributes to (\ref{dCG}) and 
$h_b(\Mism^2)\to\delta(\Mism^2-1)$. For increasing $b$, $\Mism_j$ includes more  
and more particles and therefore, with the decay of correlations, 
$h_b$ accumulates more weight at small values of $\Mism^2$ 
until $h_b(\Mism^2)\to\delta(\Mism^2)$ when $b$ becomes large. 
The transition between mostly ordered (peak of $h_b$ near $1$) and disordered (peak near $0$) 
regions happens at a value of $b$ which strongly increases with decreasing 
temperature (see Fig.~\ref{hist_mismatch.fig}). 

\vskip 0.5cm

The coarse-graining length $b$ is our means to probe 
the extent of correlations in the mismatch field on lengths up to $b$. 
Therefore, from the histogram $h_b(\Mism^2)$ we define a static correlation length $\xi_B$ 
as the coarse-graining distance $b$ for which 
the peak-location has moved from one to $1/e$. 
Our results are robust and the conclusions presented here hold also for other definitions of $\xi_B$,   
see Appendix \ref{app2}. 
Figure \ref{xi_T.fig} shows the correlation length $\xi_B$ 
as a function of temperature. 
In the high temperature liquid phase
we observe a low value of $\xi_B$ of the order 
of the particles diameter $\sigma$, weakly dependent on temperature. 
In the supercooled regime, however, $\xi_B$ increases considerably. 
For the lowest temperatures at which we can equilibrate the system,  
$T\approx 0.40$, we find $\xi_B\approx 4 \sigma$ for the 
largest systems studied here. 

\begin{figure}
\vspace*{0.5cm}
\includegraphics[width=1.0\linewidth]{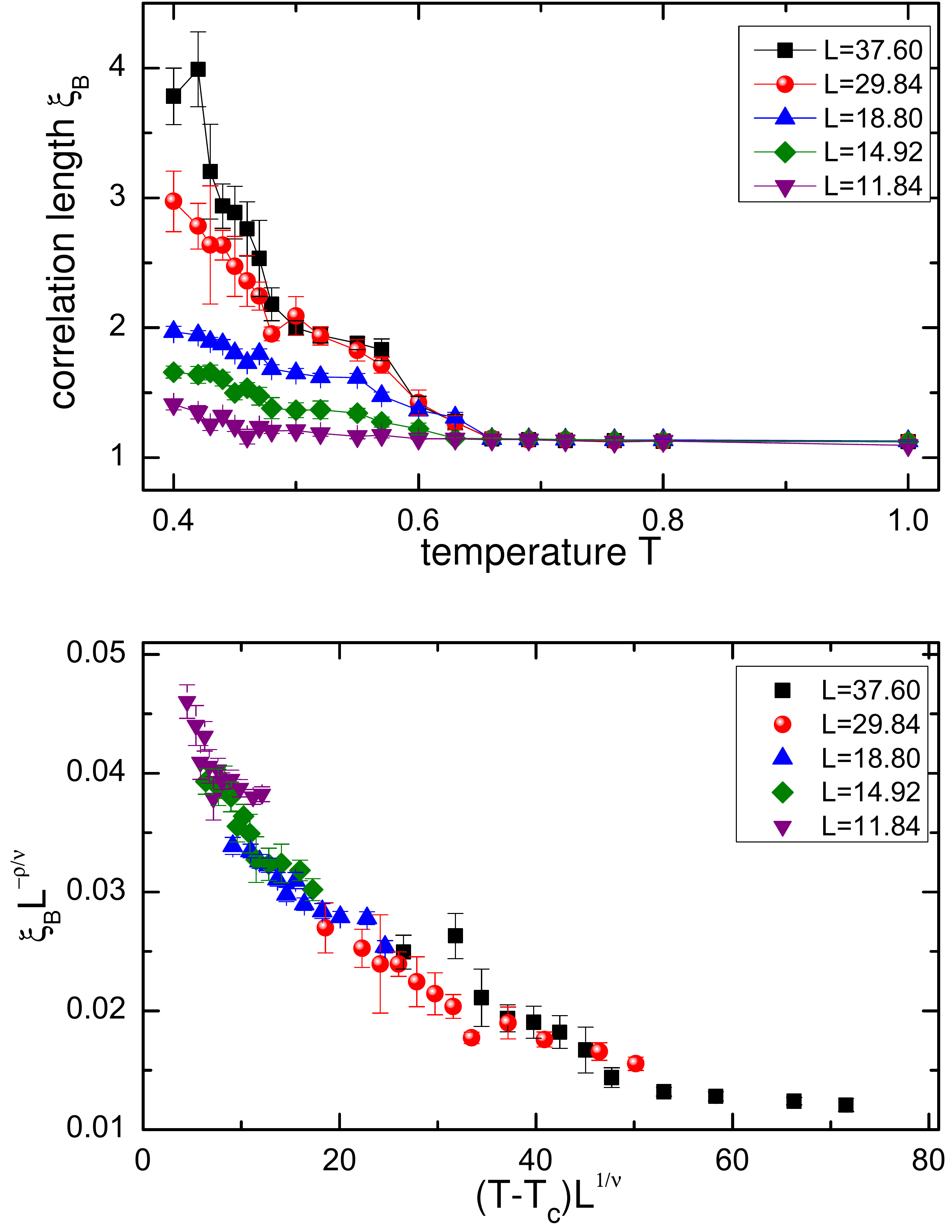}
\caption{(color online).
Top: Static correlation length $\xi_B$ of coarse-grained mismatch vectors as function of 
temperature for different system sizes. 
Bottom: Data collapse of $\xi_{B,L}(T)$ according to Eq.~(\ref{scaling}).}
\label{xi_T.fig}
\end{figure}

We also observe in Fig.~\ref{xi_T.fig} a systematic finite-size effect 
of $\xi_B(T)$ in the supercooled regime, where the correlation length 
grows with the linear size $L$ of the system. We investigate this dependence 
in more detail because, together with the significant increase of $\xi_{B}$, 
this behavior is typical of critical phenomena \cite{Barber}, 
where the correlation length $\xi$ associated with the order parameter 
characterizing the transition diverges at the critical temperature $T_c$ 
as $\xi\sim (T-T_c)^{-\nu}$. 
Therefore, we perform the same finite-size analysis which in critical phenomena
is used to extract the critical behavior from data obtained 
in a finite system. 
If the increase of $\xi_{B}$ upon lowering the temperature is indeed due to a 
diverging static correlation length $\xi$ underlying the glass transition, 
then $\xi_{B}$ should diverge as well, and in the infinite system  
$\xi_B(T)\sim t^{-\rho}$, where $t=(T-T_c)$ is the distance 
from the critical point $T_c$ and $\rho$ the critical exponent. 
On the basis of the scaling hypothesis for critical phenomena \cite{Barber}, 
the corresponding quantity $\xi_{B,L}(T)$ in a finite system of 
linear size $L$ should follow the behavior 
\begin{equation} \label{scaling}
 \xi_{B,L}(T) \sim L^{\rho/\nu} Q_{\xi_B}(L^{1/\nu}t), 
\end{equation}
where $Q_{\xi_B}(x)$ is a universal scaling function and $\nu$ is the critical
exponent associated to $\xi$.
Therefore, we plot $\xi_{B,L}(T)L^{-\rho/\nu}$ as a function of the 
scaling variable $(T-T_c)L^{1/\nu}$ in Fig.~\ref{xi_T.fig}. 
We identify the critical temperature $T_c$ with the Kauzmann temperature $T_K$,
where a phase transition due to an entropy crisis should be located according 
to RFOT and related theories \cite{wolynes89,cavagna_pedestrian}.  
For the system used here, $T_K$ has been determined numerically as 
$T_K\approx 0.30$ \cite{Sciortino_IS99,Coluzzi00}. 
Fixing this value for $T_c$, we observe that the numerical data for 
different temperatures $T$ and system sizes $L$ collapse onto a single 
master curve in agreement with Eq.~(\ref{scaling}). 
We find the best data collapse for the critical exponents 
$\rho\approx 0.9 \pm 0.1$, and $\nu\approx 0.65 \pm 0.1$, which is the case shown 
in Fig.~\ref{xi_T.fig}. 
The value of $\nu$ is very close to $\nu=2/3$, which is indeed predicted by 
RFOT for three-dimensional systems \cite{wolynes89} and to 
$\nu\approx 0.69$ deduced from light scattering experiments on o-terphenyl \cite{Donth_length}. 
The value $\nu=2/3$ is also compatible with recent simulations \cite{Tanaka_critical}, 
where, however, no proper finite-size scaling was performed.
From the quality of data collapse and the 
intersection of $\xi_B(T)L^{-\rho/\nu}$ for different system sizes \cite{Barber}, 
we estimate that the critical temperature lies in the range $0.25\leq T_c \leq 0.4$. 
Using the finite-size analysis in 
this range of temperatures we end up with the estimates for the critical 
exponents $\rho \approx 0.8 \pm 0.2$ and $\nu \approx 0.7 \pm 0.15$.   
The precision of these results is limited by the increase of $\xi_B$ being too weak 
for the smallest system sizes, which is probably also the reason why previous studies 
could extract only small, roughly temperature independent length scales from
inherent structures \cite{Heuer_ISreview}. 
Moreover, the finite-size analysis in the vicinity of the critical point 
is made extremely hard by the long equilibration times. 
In spite of these limitations, the critical region is apparently large enough to be 
felt in the accessible temperature regime.

\begin{figure}
\vspace*{0.5cm}
\includegraphics[width=1.0\linewidth]{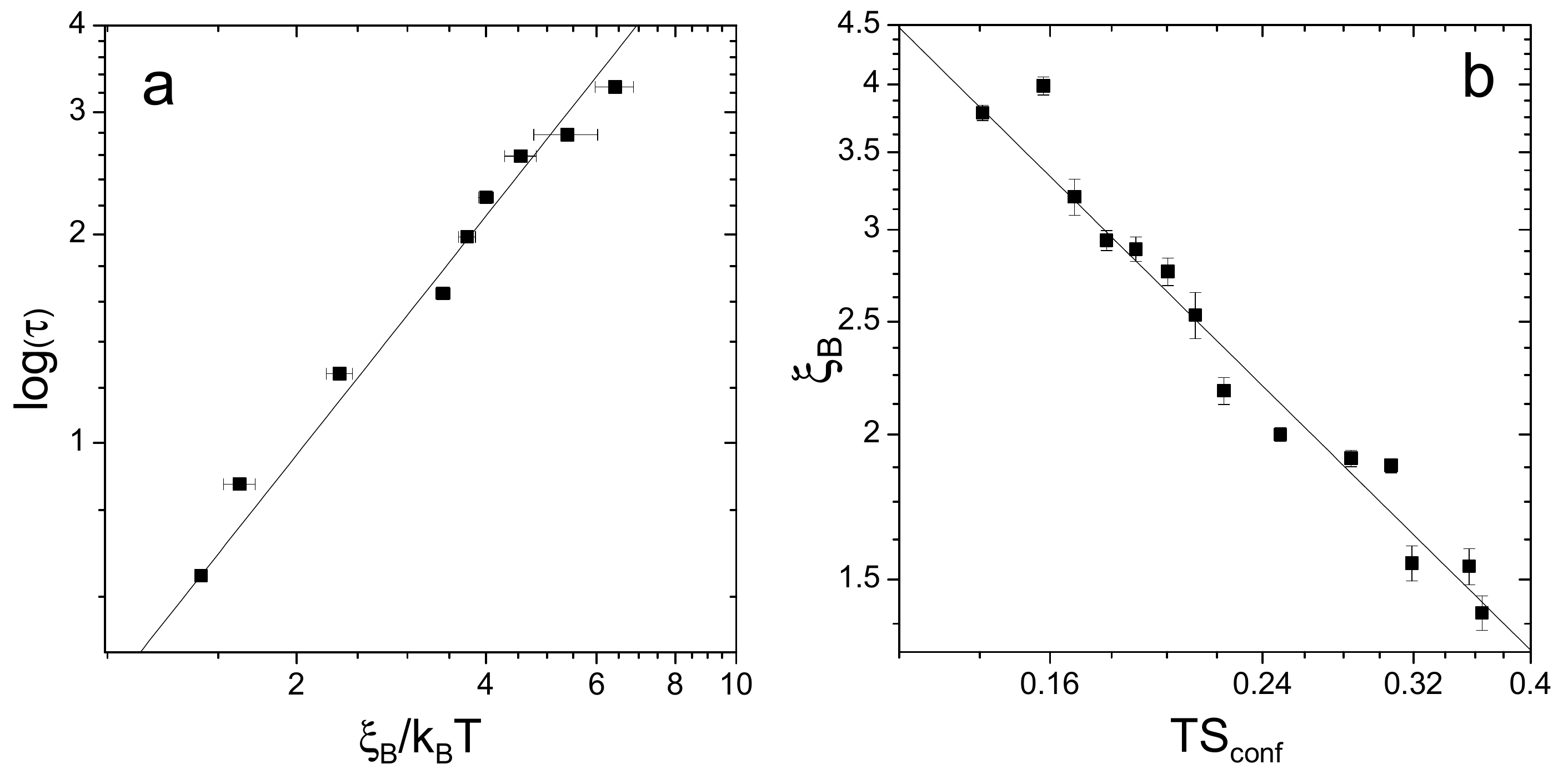}
\caption{(a) The structural relaxation time $\tau$ is plotted versus static correlation length 
$\xi_B$ for the system with $N=64000$ particles. 
The solid line is the best fit to the data and shows an exponential dependence of
the form $\tau \sim \exp{[A'(\xi_B/k_BT)^{1.14}]}$.
(b) The static correlation length is shown as a function of the configurational entropy $\sconf$. 
The solid line is the best fit to the data and shows the power-law
relation $\xi_B \sim (T\sconf)^{-1.02}$.}
\label{tau_sc.fig} 
\end{figure}

Within RFOT, the increasing relaxation time $\tau$ is linked to the free energy barrier of 
nucleating a new structure of linear size $\xi_{\rm m}$ in the liquid, 
$\tau\sim \exp{[B(T\sconf)^{-\theta/(d-\theta)}]}\sim\exp{[A(\xi_{\rm m}/k_BT)^\theta]}$  
\cite{wolynes89,XiaWolynes2000,cavagna_pedestrian}. 
The configurational entropy per particle is denoted by $\sconf$, 
$d$ is the spatial dimension and $\theta=d-1/\nu$ the exponent of the interface free energy cost of 
a nucleus.  
This gives us the possibility to further check the consistency of our data with the RFOT scenario. 
In Fig.~\ref{tau_sc.fig}-a, we plot the relaxation time $\tau$ versus 
$\xi_B/k_BT$ on a double-logarithmic scale. 
The relaxation times are large-$N$ values of $\tau$ taken from Ref.~\cite{Sastry_growinglength}, 
which have been determined therein from the final relaxation of the incoherent scattering function. 
The fitted line shows that the form 
$\tau \sim \exp{[A(\xi/k_BT)^\theta]} \sim \exp{[A'(\xi_B/k_BT)^{\theta\nu/\rho}]}$
indeed describes our numerical data quite well. The value of the exponent $\theta$ we obtain is
$\theta\approx1.6\pm0.3$. 
If we identify our correlation length
$\xi$ with the mosaic length scale $\xi_{\rm m}$ in RFOT, this value of $\theta$ we find is in agreement
with the prediction $\theta=d/2$ of RFOT \cite{wolynes89}. 
It is worth noting that our analysis is free of inconsistencies reported in 
Ref.~\cite{Sastry_growinglength} when looking at length scales from dynamical heterogeneity. 
We can also directly test the connection
between configurational entropy $\sconf$ and correlation length $\xi$, 
which according to RFOT reads 
$\xi\sim(T\sconf)^{-1/(d-\theta)}$. 
In Fig.~\ref{tau_sc.fig}-b, $\xi_B$ is plotted as a function of $T\sconf$ on a double-logarithmic scale. 
The data for the latter are taken from Ref.~\cite{Szamel_hybridMC}.
Our results indicate that the relation 
$\xi_B\sim\xi^{\frac{\rho}{\nu}}\sim(T\sconf)^{\frac{-\rho}{\nu(d-\theta)}}$ 
is indeed valid for the largest system we studied. 
For the exponent, we find 
$(d-\theta)^{-1} \approx 0.74\pm0.14$ which again is in agreement with 
the value $2/d$ predicted by RFOT and which in addition implies the 
validity of the Vogel-Fulcher-Tamman expression $\tau\sim\exp{[DT/(T-T_K)]}$ 
\cite{wolynes89}.

{\em Discussion.} -- 
Bringing together concepts from inherent structure formalism
of supercooled liquids and from deformations of amorphous solids  
offers a new perspective on the long-standing glass problem. 
In this spirit, we have employed a new method in order to elucidate growing static correlations in 
supercooled liquids.
We found direct evidence for the critical behavior of the static 
correlation length close to the glass transition, which has been speculated 
by several recent theories 
\cite{wolynes89,cavagna_pedestrian,MezardParisi2000,Procaccia_StatMechGlass}. 
In particular, our estimates of the transition temperature, critical exponents, 
as well as the link with relaxational dynamics are 
in quantitative agreement with the predictions from RFOT \cite{wolynes89}.
However further studies are needed in order to determine the numerical values more precisely.
Our results have important consequences for 
flow-induced rearrangements in colloidal glasses \cite{Schall} and 
elastic response of amorphous solids \cite{Fabien_prl}, where similar cooperative 
behavior has been observed recently.  

Our study has been performed at constant density, where 
the critical temperature $T_{c}$ extracted from $\xi_{B}$ is apparently 
below the lowest temperature at which we have been able to equilibrate 
the system. 
However, the critical point would not necessarily be located at the 
density chosen here and only a systematic investigation in the $(T,P)$-plane
will shed further light on its existence, nature, and location.  
The long-standing problem of a true thermodynamic phase transition underlying the slowing 
down of the dynamics at the glass transition can now be addressed from a different perspective, 
knowing that a direct structural signature exists.

\appendix
\section{\label{app1} Details of Numerical Simulations}

In this study, we employ the well-known Kob-Andersen binary Lennard-Jones mixture \cite{KobAndersen_1995a}, 
which is a classical model for a glass forming liquid. 
Both types of particles ($A$ and $B$) have the same unit mass and all 
particles interact with each other via a 
Lennard-Jones potential,  
\begin{equation}
 V_{ab}(r) = 4\epsilon_{ab}[(\sigma_{ab}/r)^{12}-(\sigma_{ab}/r)^6 ], 
\end{equation}
$a,b\in\{A,B\}$, where $r$ denotes the distance between particles. 
For computational convenience, 
the interactions are cut off at $r_{\rm c}=2.5\sigma_{ab}$ and shifted so that 
$V_{ab}(r_{\rm c})=0$. 
The interaction energies and diameters for the different species are given by 
$\sigma_{AA}=\sigma$, $\sigma_{AB}=0.8\sigma$, $\sigma_{BB}=0.88\sigma$, 
$\epsilon_{AA}=\epsilon$, $\epsilon_{AB}=1.5\epsilon$, and 
$\epsilon_{BB}=0.5\epsilon$. 
The composition of $A$ and $B$ particles is chosen as $80$:$20$ and the 
particle density $\rho=N/V$ is $\rho=1.2/\sigma^3$. 
Simulations are performed in the $NVT$-ensemble, for a cubic simulation box 
subject to periodic boundary conditions. 
Standard molecular dynamics simulations are performed with the 
simulation package LAMMPS \cite{lammps}, using the 
velocity Verlet integrator with a time step increasing with temperature 
from $\Delta t = 0.001\, t_{\rm ref}$ to $0.005\, t_{\rm ref}$, 
where $t_{\rm ref}=(m\sigma_0^2/\epsilon)^{1/2}$ is the reference Lennard-Jones 
time. An ensemble of statistically 
independent configurations are generated at high temperatures and cooled slowly, by 
coupling the system to a Nos\'{e}-Hoover thermostat with prescribed cooling 
protocol. The lowest temperature configurations have been equilibrated for 
$5\times10^7$ MD steps. We use dimensionless temperature $T$ measured in 
energy units $\epsilon$. For the finite-size analysis we use system 
consisting of $N=2000$, $4000$, $8000$, $32000$ and $64000$ particles, which correspond 
to system sizes of 
$L=V^{1/3}=11.8, 14.9, 18.8, 29.84$, and $37.6\,\sigma$, respectively.

We obtain the inherent structure configurations $X^{\rm q}$ from the equilibrated 
MD configurations $X$ by locally minimizing the total potential energy  
$V_{\rm tot}=\sum_{i<j}V(r_{ij})$, using the conjugate gradient method. 
We use the Polak-Ribi\`{e}re version of the algorithm as 
implemented in LAMMPS \cite{lammps}.   
The minimization is stopped when the potential energy change is less than 
a tolerance value $10^{-7}\epsilon$. We verified that the results are 
insensitive to a further decrease of the tolerance level.

\section{\label{app2} Definition of correlation length}
The correlation length $\xi_B$ can be defined from the distribution of coarse-grained
mismatch vectors in slightly different ways. 
In the manuscript, we have identified $\xi_B$ with that value of $b$, 
where the peak location of $h_b$ has moved from $1$ to $1/e$. 
Alternatively, we have also considered that from the mean-squared lengths 
$\int\! y\, h_b(y){\rm d}y = B(b)^2$
one can obtain the so-called coarse-graining function
$B(b)$, defined as the average mismatch length when smeared out over a sphere of radius $b$,
$B(b)=\ave{\Mism_j(b)^2}^{1/2}$, as in Ref.~\cite{Fabien_prl}.
By construction, $B(0)=1$ and $B\to 0$ when the coarse-graining length $b$ spans the
whole system, since there is no overall mismatch, $\sum_j\mism_j' \approx 0$.
For high temperatures, $B(b)$ decays so rapidly that we find it difficult to extract
reliable values for the correlation length.
For relatively low temperatures, we observe an exponential decay
$B(b)\propto\exp{[-b/(2\xi'_{B})]}$, with a well-defined correlation length $\xi'_{B}$.
This regime is similar to the case of amorphous solids \cite{Fabien_prl}.
We have introduced a factor $2$ in the definition of $\xi'_{B}$
for a closer similarity to the choice in Ref.~\cite{Fabien_prl},
where coarse-graining over cubic boxes of side lengths $b$ was performed.
For a sharply peaked distribution $h_{b}(y)$, $\xi_B=\xi'_B$.
In the general case, they will give slightly different values but
we have verified that the temperature dependence of $\xi'_B$ in the low
temperature regime is fully consistent with the results extracted from
Fig.~3.
Finally, we have also considered the length scale
$\xi''_{B}=b_{1/3}$, where $b_{1/3}$ is the value of $b$ where one-third of the
particles have $\Mism_j^2>1/2$ and two-thirds $\Mism_j^2<1/2$.
We expect $\xi''_{B}$ to be close to $\xi_{B}$ but
somewhat smoother and less noisy since integration can remove ambiguities in locating
the peak position. Also in this case the temperature dependence is fully
consistent with the one of $\xi_B$. 

Figure S1 shows the correlation lengths $\xi_B,\xi'_B,\xi''_B$ according to 
the above definitions as a function of temperature. 
The system size considered is $N=64000$ particles. 
All three definitions give consistent results. 

\begin{figure}
\includegraphics[width=1.0\linewidth]{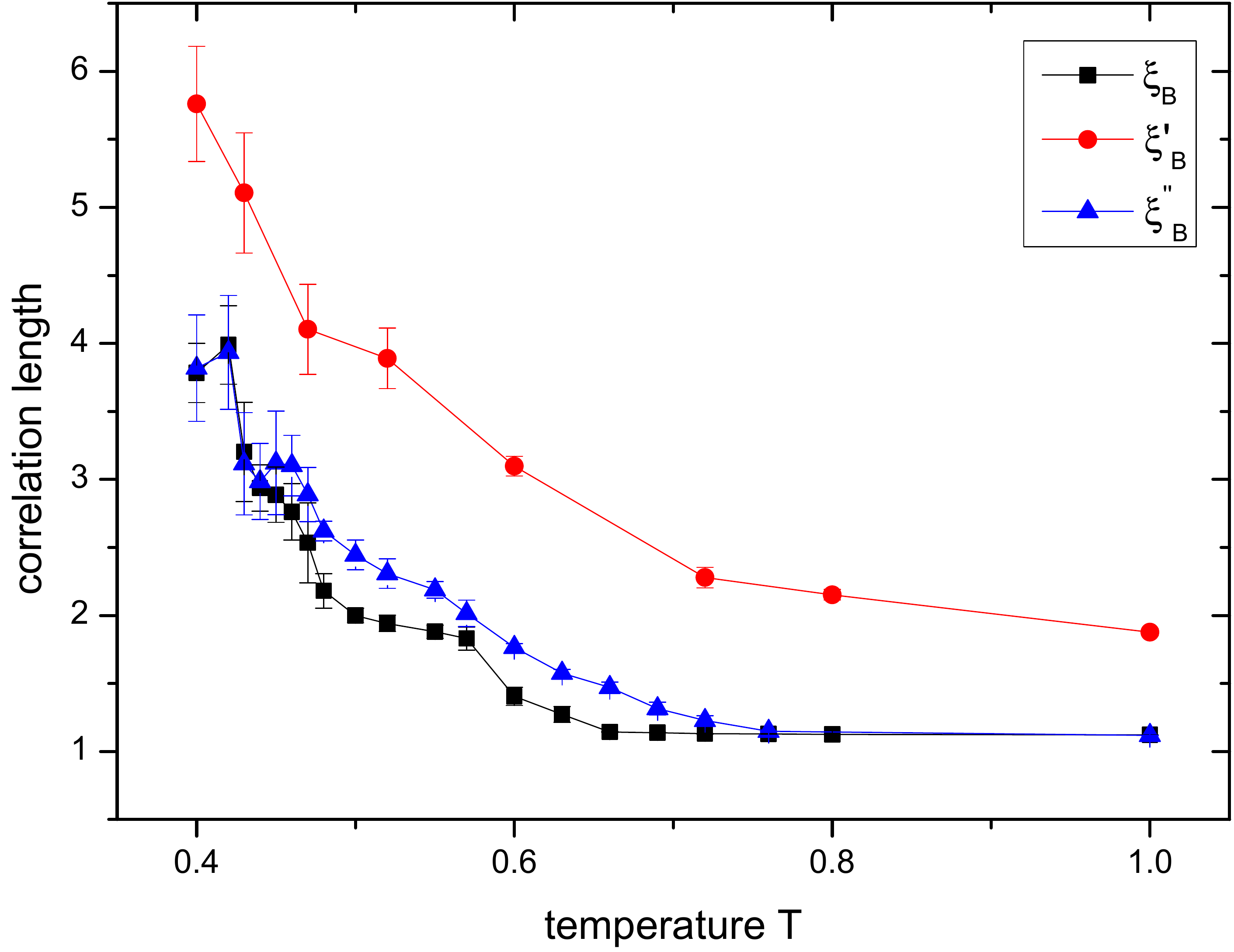}
\caption{(color online).
Correlation lengths $\xi_B,\xi'_B, \xi''_B$ defined in the text as a function of 
temperature $T$. The system contains $N=64000$ particles.} 
\end{figure}


\end{document}